\documentclass[journal,table]{IEEEtran}
\usepackage{amsmath,amsfonts}
\usepackage{algorithmic}
\usepackage{algorithm}
\usepackage{array}
\usepackage[caption=false,font=normalsize,labelfont=sf,textfont=sf]{subfig}
\usepackage{textcomp}
\usepackage{stfloats}
\usepackage{url}
\usepackage{verbatim}
\usepackage{graphicx}
\usepackage{cite}
\usepackage{orcidlink}
\usepackage{fontawesome}
\usepackage{booktabs}
\usepackage{array}
\usepackage{fancyhdr}

\definecolor{LightPink}{rgb}{1.0, 0.9, 0.9}
\definecolor{LightBlue}{rgb}{0.9, 0.95, 1.0}

\begin{document}

\title{Enhancing Resilience for IoE: A Perspective of Networking-Level Safeguard}

\author{
Guan-Yan Yang$^{\orcidlink{0009-0002-2539-9057}}$,~\IEEEmembership{Graduate Student Member,~IEEE,} 
\and
Jui-Ning Chen$^{\orcidlink{0009-0002-6508-130X}}$,
\and
Farn Wang$^{\orcidlink{0000-0002-0425-6500}}$,~\IEEEmembership{Member,~IEEE,}
\and
Kuo-Hui Yeh$^{\orcidlink{0000-0003-0598-761X}}$,~\IEEEmembership{Senior Member,~IEEE}


\thanks{This work was partially supported by the Taiwan Academic Cybersecurity Center at the National Taiwan University of Science and Technology and by the National Science and Technology Council (NSTC) under Grants 114-2221-E-002-217, 114-2622-E-A49-022, 114-2221-E-A49-210, 114-2634-F-011-002-MBK, 114-2923-E-194-001-MY3, and MOST 110-2221-E-002-069-MY3. Additional financial support was provided by National Taiwan University (NTU) and the NTU Core Consortium Project as part of the Higher Education Sprout Project by the Ministry of Education in Taiwan, under Grants NTU-CC-114L895501 and NTU-G0647.

Further partial financial support was provided by the Department of Industrial Technology, Ministry of Economic Affairs, under the "2025 ITRI Advanced Research Program" (Grant No.: 114-EC-17-A-21-0337) and by the Hon Hai  Research Institute, Taipei Taiwan (Project No.: 114UA90042).

The authors would like to express their gratitude for the financial support.  

\textit{(Corresponding author: Kuo-Hui Yeh \& Farn Wang.)}}
\thanks{Guan-Yan Yang and Farn Wang are with the Department of Electrical Engineering at National Taiwan University, Taipei 106319, Taiwan R.O.C. 
(e-mail: f11921091@ntu.edu.tw; farn@ntu.edu.tw).
}
\thanks{Jui-Ning Chen is with the Institute of Information Science, Academia Sinica, Taipei, Taiwan,
and also with Google.
(e-mail: luana880910@iis.sinica.edu.tw)}
\thanks{Kuo-Hui Yeh is with the Institute of Artificial Intelligence Innovation, National Yang Ming Chiao Tung University, No. 1001, Da Hsueh Road, East District, Hsinchu
City, 300093, Taiwan R.O.C., 
and also with the Department of Information Management, National Dong Hwa University, No. 1, Sec. 2, Da Hsueh Road, Shoufeng, Hualien, 974301, Taiwan R.O.C.
(e-mail: khyeh@nycu.edu.tw).}
}



\maketitle

\thispagestyle{fancy}
\fancyhf{}
\fancyhead[L]{\footnotesize
    This is the author's version that has been accepted for publication in IEEE Network. The final version will be available at IEEE Xplore once published. \\
    © 2025 IEEE. Personal use of this material is permitted.  Permission from IEEE must be obtained for all other uses, in any current or future media, including reprinting/republishing this material for advertising or promotional purposes, creating new collective works, for resale or redistribution to servers or lists, or reuse of any copyrighted component of this work in other works.
}
\renewcommand{\headrulewidth}{0pt}

\begin{abstract}
The Internet of Energy (IoE) integrates IoT-driven digital communication with power grids to enable efficient and sustainable energy systems. Still, its interconnectivity exposes critical infrastructure to sophisticated cyber threats, including adversarial attacks designed to bypass traditional safeguards. Unlike general IoT risks, IoE threats have heightened public safety consequences, demanding resilient solutions. From the networking-level safeguard perspective, we propose a Graph Structure Learning (GSL)-based safeguards framework that jointly optimizes graph topology and node representations to resist adversarial network model manipulation inherently. Through a conceptual overview, architectural discussion, and case study on a security dataset, we demonstrate GSL's superior robustness over representative methods, offering practitioners a viable path to secure IoE networks against evolving attacks. This work highlights the potential of GSL to enhance the resilience and reliability of future IoE networks for practitioners managing critical infrastructure. Lastly, we identify key open challenges and propose future research directions in this novel research area.
\end{abstract}


\section{Introduction}
\label{sec:introduction}
\IEEEPARstart{T}{he} Internet of Energy (IoE) signifies a transformative evolution of traditional energy systems by integrating energy infrastructure with the Internet of Things (IoT). This convergence enables intelligent energy generation, transmission, and consumption, bolstering efficiency and sustainability~\cite{6246751}. At the heart of this transformation lies the smart grid, where real-time data from smart meters, sensors, actuators, and distributed energy resources enables dynamic control and optimization. Advanced Metering Infrastructure (AMI) often underpins this communication, facilitating bi-directional data exchange between consumers and utilities~\cite{8676005}.

However, the same connectivity that empowers IoE also exposes critical vulnerabilities. Each connected device represents a potential attack vector, drastically expanding the attack surface of energy infrastructure. While conventional threats such as Denial-of-Service (DoS) and data theft persist, a new class of threats—\textit{adversarial attacks}—is emerging~\cite{10403639}. These adversarial attacks exploit security mechanisms themselves rather than directly disrupting operations. For instance, attackers may manipulate sensor data or network traffic to mislead AI-based Intrusion Detection Systems (IDS), causing malicious activities to be misclassified as benign, or generating false alarms that effectively disable network defenses. Incidents like the Ukrainian power grid attack~\cite{10633827} demonstrate the catastrophic consequences of exploiting IoT vulnerabilities in energy networks.

Traditional networking-level defenses, such as firewalls, signature-based IDS, and machine learning-based anomaly detectors, are often ill-equipped to counter such sophisticated and adaptive attacks. These systems rely on static assumptions about network structure or data distribution, which adversaries can manipulate. The need for resilient, adaptive security solutions specifically tailored to the dynamic and heterogeneous nature of IoE networks is critical.

This article offers a tutorial perspective on enhancing IoE resilience at the networking level. We introduce Graph Structure Learning (GSL) ~\cite{gsl_survey_ref} as a promising approach for safeguarding IoE networking and propose a GSL-based framework for integrating systems like AMI. GSL's ability to simultaneously learn network structure and data representations enhances robustness against adversarial manipulation, making it particularly suitable for IoE's evolving threat landscape.

Our key contributions are:
\begin{itemize}
    \item A brief overview of the IoE networks landscape, its inherent threats, and the limitations of existing networking-level safeguard methods.
    \item A conceptual introduction to GSL and its application illustration as a networking-level safeguard for IoE environments.
    \item A demonstration of how GSL's co-optimization of graph structure and node representations enhances resilience for IoE against adversarial attacks.
    \item A case study for IoE using a widely recognized dataset to validate the proposed GSL-based networking-level safeguard framework under adversarial conditions in the network model.
\end{itemize}

\begin{figure*}[th]
    \centering
    \includegraphics[width=1.0\textwidth]{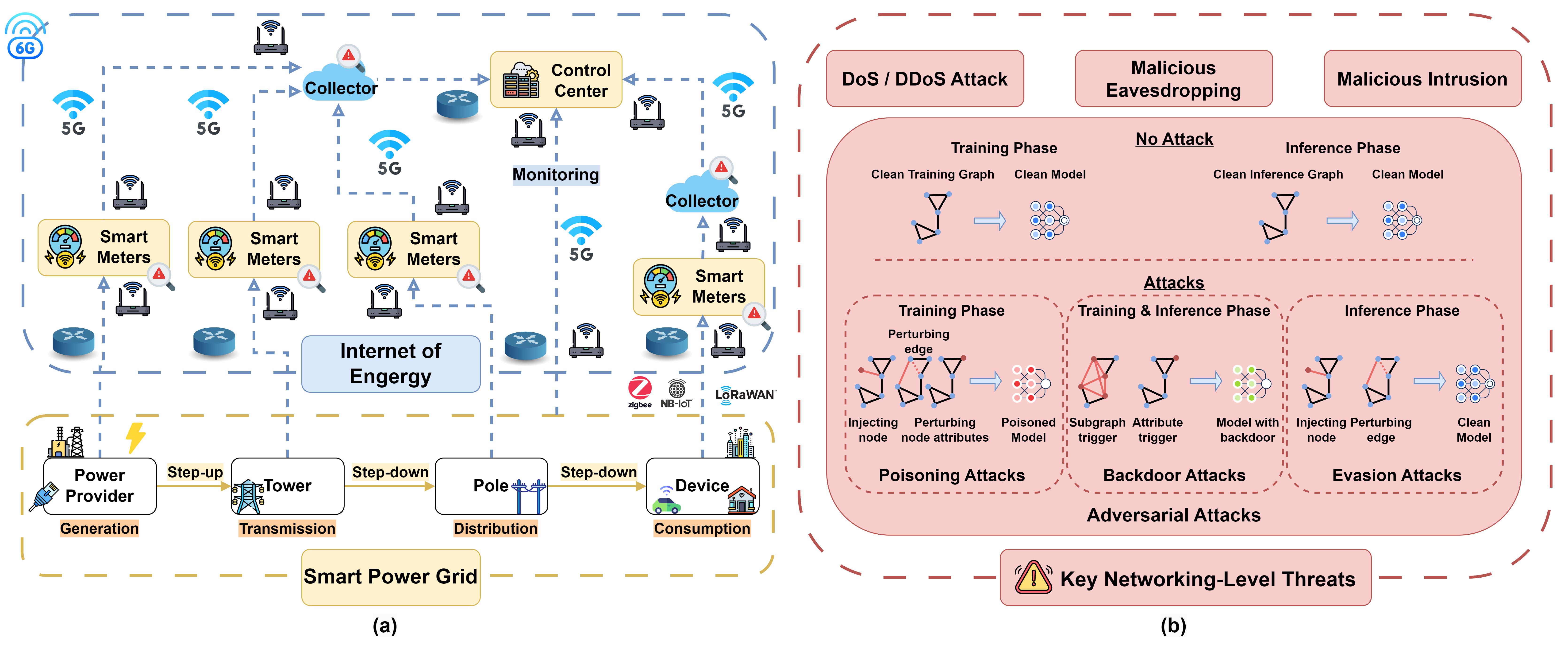}
        \caption{
    Conceptual overview of the Internet of Energy ecosystem and associated networking-level security threats. \\ 
    \textbf{(a)} Illustrates the interconnected architecture of a modern smart power grid, showing the flow from power generation, through transmission and distribution infrastructure, to end-consumer devices. Key components like smart meters, data collectors, and control centers are depicted, linked by diverse communication technologies (such as 5G, LoRaWAN, NB-IoT, Zigbee) forming the \textbf{Internet of Energy}. This highlights the extensive digital connectivity overlaying the physical power system. \\
    \textbf{(b)} Highlight major networking-level security threats targeting the IoE. It distinguishes between conventional attacks such as DoS/DDoS, Malicious Eavesdropping, and Malicious Intrusion, and the more advanced Adversarial Attacks aimed explicitly at ML systems used within the IoE. The diagrams for Adversarial Attacks (Poisoning, Backdoor, Evasion) schematically show how attackers can manipulate data or ML models during training or inference phases to compromise system integrity or bypass security measures.
    }
    \label{fig:IoE_Network_Concept}
    \vspace{-3mm}
\end{figure*}

\section{A Brief Overview of the IoE Network and Its Networking-Level Threats}
\label{sec:landscape}

To protect the Internet of Energy (IoE) effectively, we must first understand its environment, structure, and inherent vulnerabilities. The IoE fundamentally merges Operational Technology (OT), which controls physical energy hardware (generators, transformers, and switches), with Information Technology (IT), which manages data through networks, servers, and software. This powerful integration aims to create a smarter, more efficient energy grid. Key objectives include improved integration of renewable energy, optimized power flow, predictive maintenance capabilities, and support for new consumer services like demand-response programs.

This convergence of OT and IT permeates the entire energy value chain, broadly categorized into interconnected domains, as visualized in the (a) of Figure~\ref{fig:IoE_Network_Concept}.

Connecting these diverse components across vast geographical areas requires a complex tapestry of communication technologies, also depicted in Figure~\ref{fig:IoE_Network_Concept}. Wired solutions like power line communication cleverly utilize the existing electrical wiring for data transfer, and traditional Ethernet is often used where feasible. However, the sheer scale and physical distribution of IoE devices necessitate extensive use of wireless technologies. These range from high-bandwidth options like Wi-Fi and cellular (including 5G and future 6G, offering low latency and high capacity) to low-power, wide-area networks (LPWANs) such as LoRaWAN and NB-IoT, which are ideal for connecting numerous low-data-rate devices like sensors and meters over long distances with minimal power consumption~\cite{6246751}. The AMI system alone constitutes a massive network, potentially linking millions of smart meters through various collector nodes and gateways to utility data centers and control centers, which oversee grid operations.

While this hyper-connectivity unlocks unprecedented grid modernization and efficiency capabilities, it fundamentally changes the security landscape. Every connected device, sensor, communication link, and control system becomes a potential entry point or target for malicious actors. This expanded attack surface introduces significant networking-level vulnerabilities that could compromise grid stability, data integrity, and consumer privacy. The (b) of Figure~\ref{fig:IoE_Network_Concept} provides a taxonomy of these key threats:

\begin{enumerate}
    \item \textbf{Denial-of-Service (DoS) / Distributed Denial-of-Service (DDoS) Attacks:} The IoE depends on reliable communication for monitoring and control. Attackers can target this by overwhelming network links or control systems with useless traffic. They might compromise many IoE devices (which often have limited processing power) and use them together (as a 'botnet') to launch a large-scale DDoS attack. This flood of traffic can block legitimate communication, preventing operators from seeing grid conditions or sending vital control commands, potentially causing outages or instability.
    \item \textbf{Malicious Intrusion and Data Manipulation:} If an attacker gains access to the IoE network or compromises a device, they can interfere with operations. This could mean changing smart meter readings to steal energy or disrupt billing. More seriously, attackers could send false commands to equipment in substations, potentially damaging hardware or causing blackouts, or alter sensor data to give operators a false picture of the grid's status.
    \item \textbf{Malicious Eavesdropping:} Data transmitted across the IoE, especially using wireless or older protocols, may not be adequately encrypted. Attackers who intercept these communications could steal sensitive operational details (such as power flow information or equipment status) or private customer data (like energy usage patterns). This stolen information could be used for corporate espionage, planning further attacks, or violating privacy regulations.
    \item \textbf{Adversarial Attacks Against Smart IoE Systems~\cite{10477407}:} As IoE systems become more intelligent, they increasingly incorporate machine learning (ML) algorithms for crucial tasks like anomaly detection, intrusion prevention, load forecasting, and optimizing power flow. Adversarial attacks are designed to deceive these ML models, undermining the 'smart' defenses and operational logic. These attacks exploit how ML models learn from data and make predictions. As illustrated schematically in Figure~\ref{fig:IoE_Network_Concept}, key types include:
    \begin{itemize}
        \item \textit{Poisoning Attacks:} These occur during the ML model's training phase. The attacker subtly injects carefully crafted malicious data into the dataset used to train the model, perturbs the edges and node attributes, or injects nodes. The goal is to degrade the model's performance on legitimate tasks (such as making a detection system less likely to spot real attacks).
        \item \textit{Evasion Attacks:} These happen during the model's operational (inference) phase after training. The attacker crafts malicious inputs (such as network packets and sensor readings) intentionally designed to be misclassified as benign by the ML model. This allows attacks to slip past ML-based security filters undetected.
        \item \textit{Backdoor Attacks:} While sometimes initiated through poisoning, backdoor attacks focus on creating a specific, hidden vulnerability. The attacker manipulates the model (or the data patterns it learns, such as particular network subgraph structures) to behave normally most of the time but responds incorrectly or grants unauthorized access when presented with a secret trigger known only to the attacker.
    \end{itemize}
\end{enumerate}

The potential impact of adversarial attacks is particularly alarming, as they target the intelligence designed to enhance grid reliability and security~\cite{10403639}. A compromised ML system could fail to detect genuine threats or initiate incorrect control actions based on manipulated inputs, potentially leading to cascading failures or widespread blackouts. Consequently, securing the IoE effectively demands a paradigm shift from static, perimeter-based security to more sophisticated, adaptive defense mechanisms. These solutions must detect not only known attack patterns but also subtle manipulations of data and system behavior inherent in adversarial attacks, ensuring the overall resilience of the energy infrastructure.

\section{Strengths and Limitations of Current Networking-Level Safeguards}
\label{sec:limitations}
The IoE represents the convergence of modern energy systems with advanced networking technologies, enabling real-time monitoring, decentralized energy trading, and intelligent, autonomous control. As the IoE ecosystem expands, it simultaneously broadens the potential attack surface, necessitating the development of robust, adaptive, and scalable security mechanisms. While IoE networks employ various security mechanisms at the networking level to monitor and control traffic flows, their effectiveness against sophisticated adversarial threats remains a critical area of research.

In recent years, there has been a surge in research aimed at safeguarding IoE infrastructures through integrated and decentralized frameworks. Table~\ref{tab:comparison} summarizes key approaches potentially for IoE networking-level safeguards, highlighting their respective ability for networking-level security, adversarial attacks, scalability, and main contribution.

\begin{table*}[ht]
\caption{Comparison with existing studies about network resilience.}
\label{tab:comparison}
\centering
\renewcommand{\arraystretch}{1.2}
\begin{tabular}{
    |>{\columncolor{LightBlue}}m{1.7cm}
    |m{0.7cm}
    |m{1.1cm}
    |m{1.8cm}
    |m{1.8cm}
    |m{1.5cm}
    |m{4.9cm}|
}
\hline
\rowcolor{LightPink}
\textbf{Study} & \textbf{Year} & \textbf{Model} & \textbf{Networking-Level Security} & \textbf{Resilience to Adv. Attacks \footnotemark} & \textbf{Scalability} & \textbf{Main Contribution} \\
\hline
Song \textit{et al.} ~\cite{song2020real} & 2020 & ResNet + ALSTM & {\large \faCheckCircle} & {\large \faTimesCircle} & {\large \faTimesCircle} & Develops an intrusion detection system combining ResNet and Attention-based LSTM (ALSTM) for enhanced detection accuracy in IoE environments. \\
\hline
Lee \textit{et al.} ~\cite{lee2020deep} & 2020 & DNN & {\large \faCheckCircle} & {\large \faTimesCircle} & {\large \faTimesCircle} & Implements a deep learning approach to detect various cyber-attacks within AMI networks. \\
\hline
Wu \textit{et al.} ~\cite{wu2021graph} & 2022 & GNN & {\large \faTimesCircle} & {\large \faTimesCircle} & {\large \faCheckCircle} & Applies GNNs to identify anomalies in industrial IoT settings, enhancing detection capabilities. \\
\hline
Su{\' a}rez-Varela \textit{et al.} ~\cite{suarez2023graph} & 2023 & GNN & {\large \faTimesCircle} & {\large \faTimesCircle} & {\large \faCheckCircle} & Provides a comprehensive overview of GNN applications in communication networks, highlighting potential use cases. \\
\hline
Luo \textit{et al.} ~\cite{10925363} & 2025 & GNN & {\large \faTimesCircle} & {\large \faTimesCircle} & {\large \faCheckCircle} & Explores the use of GNNs for evaluating trustworthiness within network entities. \\
\hline
Li \textit{et al.} ~\cite{li2025achieving} & 2025 & GNN + DRL & {\large \faCheckCircle} & {\large \faTimesCircle} & {\large \faCheckCircle} & Combines GNNs with deep reinforcement learning to bolster network resilience against various threats. \\
\hline
This paper & -- & GSL & {\large \faCheckCircle} & {\large \faCheckCircle} & {\large \faCheckCircle} & Substitute GNN to GSL for enhancing network resilience for IoE against various future networking-level threats. \\
\hline
\end{tabular}
\end{table*}

\footnotetext{Adv. Attacks: Adversarial Attacks}

Several distinct strategies have emerged. 
Deep learning-based intrusion detection systems (IDS), such as the one by Song \textit{et al.}~\cite{song2020real}, utilize ResNet and Attention-based LSTM (ALSTM) architectures to identify temporally evolving attacks in IoE environments. While this approach offers high detection accuracy, it suffers from limited scalability and high computational overhead. Similarly, Lee \textit{et al.}~\cite{lee2020deep} implemented a DNN-based IDS for AMI networks, demonstrating adaptability in dynamic environments but lacking robustness against adversarial attacks and facing challenges in scalability.

Graph-based models have shown increasing promise. Wu \textit{et al.}~\cite{wu2021graph} applied Graph Neural Networks (GNNs) to identify anomalies in Industrial IoT networks, achieving better scalability and detection performance. However, their approach, like other GNN-based systems, remains susceptible to adversarial manipulations and often demands substantial resources. Su{\' a}rez-Varela \textit{et al.}~\cite{suarez2023graph} provided a broad survey on GNN applications in communication networks, but did not directly address networking-level security. Luo \textit{et al.}~\cite{10925363} evaluated GNNs for trustworthiness within network entities, offering insights into resilience but still confronting limitations related to adversarial robustness.

The integration of GNNs with Deep Reinforcement Learning (DRL), as proposed by Li \textit{et al.}~\cite{li2025achieving}, marks a further step toward adaptive and intelligent network resilience. Their system provides strong scalability, yet this comes at the cost of high computational requirements and extensive training data needs. Moreover, both GNNs and DRLs are challenging in resisting emerging adversarial threats.

To address these shortcomings, we propose replacing GNNs with GSL methods. As outlined in Table~\ref{tab:comparison}, this shift aims to improve resilience against emerging and complex networking-level threats by jointly enabling the model to learn optimal graph structures and predictive tasks. GSL reduces vulnerability to adversarial perturbations and offers better adaptability to dynamic network environments while maintaining or enhancing scalability.


\section{Enhancing Resilience for IoE: GSL-Based Networking-Level Safeguard}
\label{sec:gsl_safeguard}

The Internet of Energy (IoE) connects everything from power grids and smart meters to electric vehicles and intelligent buildings. This vast network promises efficiency and innovation but presents significant security challenges. Traditional security tools often struggle with the dynamic nature and sheer scale of IoE systems. Adversaries can exploit vulnerabilities not just by attacking individual devices, but by manipulating the perceived structure of the network itself. This section introduces an advanced safeguard approach using Graph Structure Learning (GSL). Instead of reacting to known threats, GSL proactively learns the actual state and relationships within the IoE network, making it significantly more resilient to sophisticated attacks. This method operates at the networking level, analyzing communication patterns and device interactions to build a robust defense.

\subsection{Why Graph Structure Learning is Needed}
Many modern security tools use GNNs to model the IoE as a graph of devices (nodes) and communication links (edges) \cite{suarez2023graph}. However, GNNs have a critical weakness: they assume the provided graph is correct. In reality, IoE network data can be noisy, incomplete, or, worse, deliberately manipulated by an attacker who adds fake connections or hides real ones to mislead the security system.
Their strength lies in leveraging the connections between devices to make more informed predictions. However, this strength is also their Achilles' heel: GNNs fundamentally assume that the input graph structure is a faithful representation of reality. In the high-stakes, adversarial landscape of IoE security, this trust is a critical vulnerability. An attacker can poison the training data or craft evasive inputs by adding or removing network links, effectively manipulating the GNN's "worldview" to hide malicious activity or trigger false alarms.

\begin{figure*}[t]
    \centering
    \includegraphics[width=\textwidth]{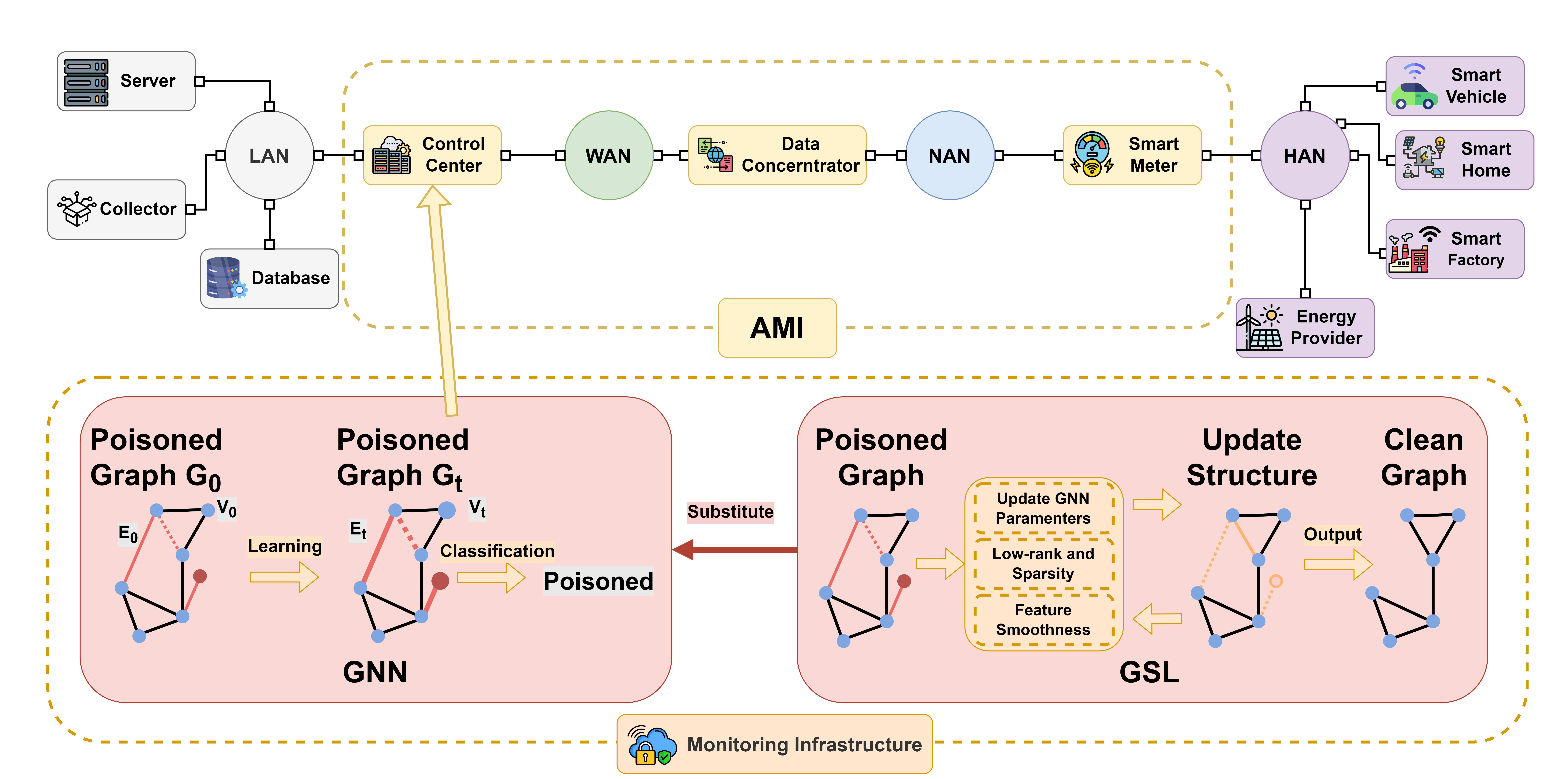} 
    \caption{Comparing GSL to GNN for network security. Data from IoE sources like Advanced Metering Infrastructure (AMI) can be noisy or deliberately poisoned by attackers. A standard GNN (left) learns directly from this potentially flawed graph, making it vulnerable to errors and manipulation. In contrast, the GSL approach (right) takes the poisoned graph data but does not trust it implicitly. It uses an iterative optimization process—guided by principles like feature smoothness and structural priors—to learn the likely true underlying structure, effectively removing spurious links and reinforcing valid ones. This results in a cleaner graph, more robust device representations, and superior resilience against misleading network data.}
    \label{fig:GSL_Concept}
\end{figure*}

Graph Structure Learning (GSL) provides a robust defense by abandoning this assumption of trust. Instead, it operates on a principle of "distrust and verify." A GSL-based system treats the initial network graph not as ground truth, but as a noisy, potentially compromised observation that must be critically evaluated and refined. As shown in Figure \ref{fig:GSL_Concept}, GSL actively "cleans" this graph by jointly learning the GNN model's parameters while simultaneously inferring the true underlying network topology. This philosophy is mathematically embodied in a unified optimization objective, which can be expressed as 
$
\min_{\mathbf{S}, \mathbf{\Theta}} \mathcal{L}_{task}(\mathbf{S}, \mathbf{X}; \mathbf{\Theta}) + \alpha \mathcal{R}_{struct}(\mathbf{S}, \mathbf{X}) + \beta \mathcal{R}_{feat}(\mathbf{S})
$.
Here, the goal is to find the optimal graph structure $\mathbf{S}$ and GNN parameters $\mathbf{\Theta}$ that minimize a combination of three terms, balanced by hyperparameters $\alpha$ and $\beta$. While the first term, $\mathcal{L}_{task}$, is the standard loss for the security task (such as correctly classifying nodes as malicious or benign), the resilience of GSL is derived from the two regularization terms.

First, the \textbf{Low-rank and Sparsity} regularizer, denoted as $\mathcal{R}_{struct}$, introduces structural priors to enforce a physically realistic network topology. The low-rank property, in turn, facilitates the discovery of community structures, allowing the model to group highly correlated devices—such as all assets connected to one substation—into a unified functional block. Sparsity is based on the principle of local communication; a smart meter, for instance, interacts exclusively with its designated data concentrator rather than with all meters across the network. This constraint is crucial for eliminating erroneous or malicious connections that would otherwise obscure the true graph structure. 

Second, the \textbf{Feature Smoothness} regularizer, denoted as $\mathcal{R}_{feat}$, enforces the realistic expectation that connected devices in the grid should behave similarly. For example, a set of Phasor Measurement Units (PMUs) on the same transmission line should report correlated phase angles. This term mathematically penalizes proposed connections between nodes with highly dissimilar feature vectors ($\mathbf{X}$). It thwarts adversarial attempts to camouflage a compromised device by falsely linking it to a high-reputation but behaviorally distinct part of the network, such as a central control server.

\begin{figure*}[!t]
    \centering
    \includegraphics[width=.95\textwidth]{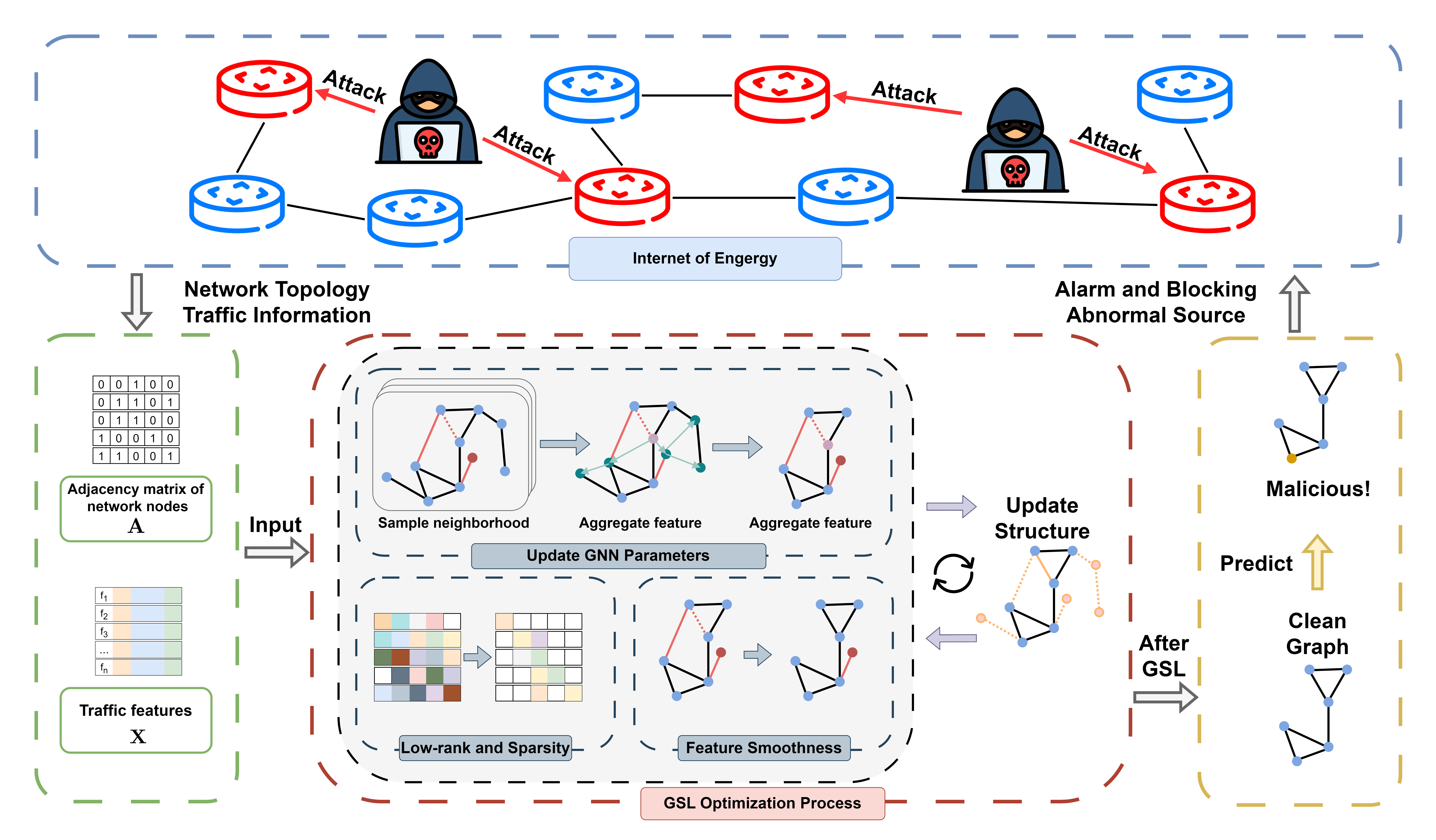} 
    \caption{The GSL optimization process. The framework takes an initial adjacency matrix ($\mathbf{A}$) and node feature matrix ($\mathbf{X}$) as input from the IoE network, which may be under attack. It then enters an iterative loop where two processes occur in tandem: (1) The GNN model updates its parameters via neighborhood sampling and feature aggregation based on the current graph. (2) The graph structure itself is optimized, guided by priors such as feature smoothness, low-rankness, and sparsity. The newly refined structure is then fed back into the GNN for the next iteration. This cycle repeats, leading to a clean, robust graph that allows for accurate prediction and the effective identification of malicious nodes.}
    \label{fig:IoE_gsl_optimization}
\end{figure*}

Crucially, this optimization is not a one-shot calculation but an \textbf{iterative refinement process}, as illustrated in Figure \ref{fig:IoE_gsl_optimization}. The framework enters a loop where it first updates the GNN's parameters based on the current version of the graph, and then uses these newly learned node representations to update and improve the graph structure itself. By cycling through this process, GSL methodologies, such as the influential ProGNN \cite{prognn}, systematically filter out deception and statistical noise, allowing the model to converge on a clean, robust graph that reflects the true functional relationships within the IoE infrastructure. In this work, we integrate the GSL optimization process in \cite{prognn} into our framework.

\subsection{Framework for GSL-Based Networking-Level Safeguard}

\begin{figure*}[t]
    \centering
    \includegraphics[width=\textwidth]{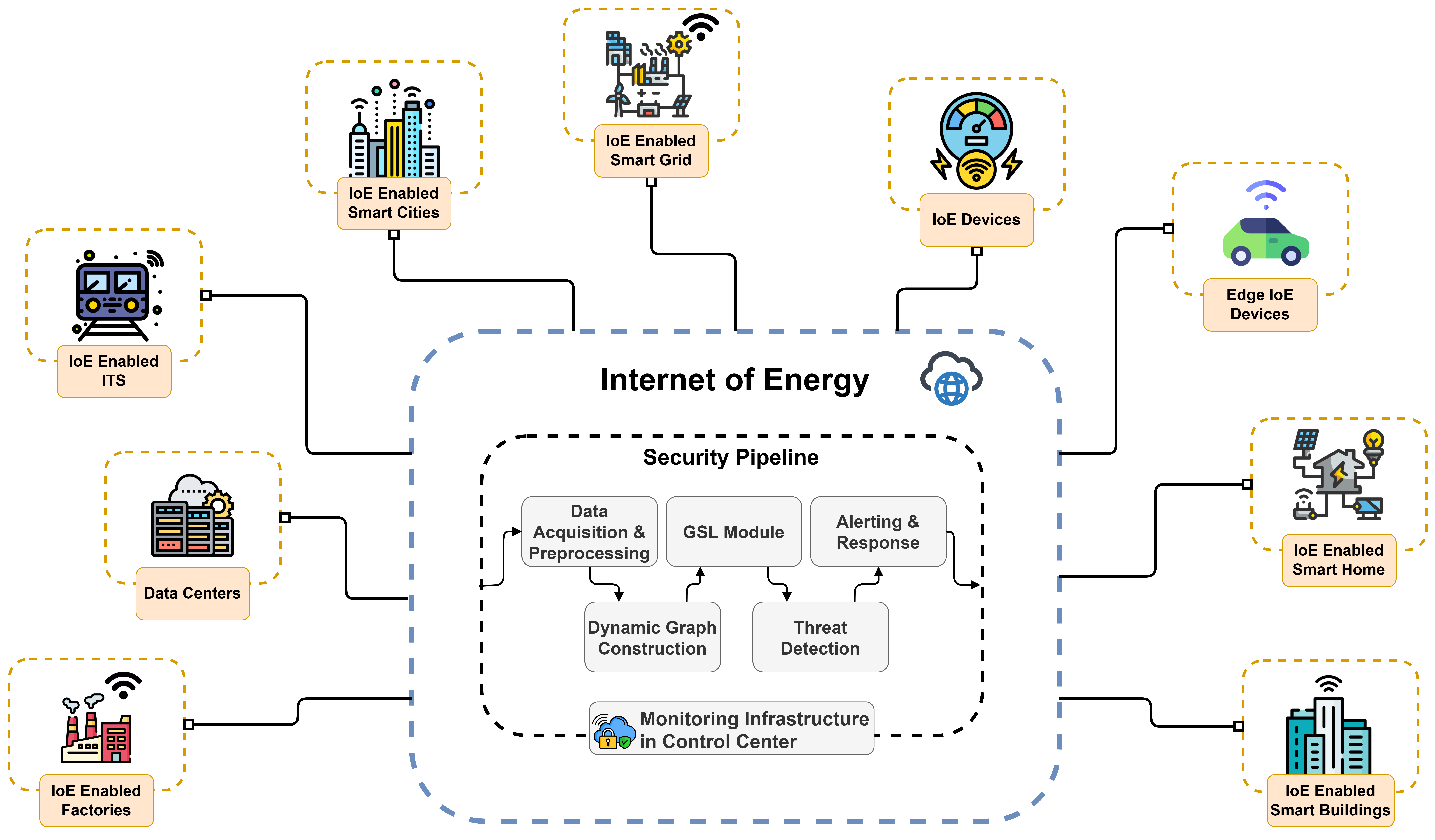} 
    \caption{Architecture of the proposed GSL-based networking-level safeguard. The diagram shows the interconnected Internet of Energy landscape, including diverse components like smart grids, buildings, homes, electric vehicles, and data centers. The \textbf{Security Pipeline} (inner dashed box) is central to the security approach. This pipeline represents a workflow that continuously processes network data: It starts with \textit{Data Acquisition \& Preprocessing} from various IoE sources and monitoring infrastructure. \textit{Dynamic Graph Construction} builds time-based snapshots of the network's state. The core \textit{GSL Module} then intelligently refines this network view, learning the true structure and device profiles. \textit{Threat Detection} uses this refined understanding to spot anomalies or attacks. Finally, \textit{Alerting \& Response} notifies administrators or triggers automated defenses. This entire pipeline helps secure the interconnected IoE environment.}
    \label{fig:framework_arch}
    \vspace{-3mm} 
\end{figure*}

To apply GSL for IoE security, we propose a framework that acts as a continuous security monitoring pipeline. This system operates at the networking level, integrating seamlessly with existing monitoring infrastructure. Figure~\ref{fig:framework_arch} provides a high-level view of this framework, showing how the core \textbf{Security Pipeline} fits within the larger Internet of Energy ecosystem, encompassing elements like smart grids, charging stations, intelligent buildings, and data centers, drawing information from monitoring systems and control centers.

The Security Pipeline involves several key stages, operating continuously:

\begin{enumerate}
    \item \textbf{Gathering Network Intelligence (Data Acquisition \& Preprocessing):} The system first collects diverse data from across the IoE network. This includes information like which devices are communicating (traffic logs), device types, status updates, and connectivity reports from sources such as smart meters (AMI), network switches, routers, and central management systems. This raw data is formatted for analysis.
    \item \textbf{Creating Network Traffic Snapshots (Dynamic Graph Construction):} Because IoE networks change constantly, the system does not rely on a single static network traffic graph. Instead, it uses the preprocessed data to build dynamic snapshots – essentially, drawing a network graph showing connected devices (nodes), their communication links (edges), and their current traffic features, representing the network's state at specific points in time or over short intervals. For example, it might construct a new graph every five minutes, where nodes are devices and edges represent communication.
    Node features can be aggregated statistics over this time window, such as packet counts, protocol types, and data volume.
    \item \textbf{Graph Refinement (GSL Module):} This is the core engine of the framework. It takes the potentially noisy or incomplete network snapshots and applies GSL techniques. It analyzes patterns, device behaviors, and connection likelihoods to infer the most probable \textit{true} network structure, filtering out noise and correcting inconsistencies. Simultaneously, it generates rich, context-aware digital fingerprints (node embeddings) for each device, reflecting its characteristics and verified role within the refined network structure.
    \item \textbf{Spotting Suspicious Activity (Threat Detection):} Using the refined graph and robust device embeddings from the GSL module, this stage identifies anomalies. It can detect threats like a device communicating in an uncharacteristic way, the emergence of an isolated cluster of suspicious devices, or patterns that match known attack signatures.
    \item \textbf{Raising the Alarm (Alerting \& Response):} Once a potential threat is identified with sufficient confidence, the system takes action. This might involve sending detailed alerts to security personnel for investigation or automatically triggering pre-defined responses, such as isolating a suspicious device from the network or blocking malicious traffic, to contain the threat quickly.
\end{enumerate}

\section{Case Study: Evaluating GSL-Based Networking-Level Safeguard for IoE Networks}
\label{sec:case_study}

To validate the effectiveness of our GSL-based safeguard framework, we conducted experiments simulating an IoE network under various adversarial conditions.

\subsection{Experimental Setup}
We simulated an IoE network environment monitored by our safeguard, focusing on an AMI use case. The framework was implemented in Python 3.8.20 using PyTorch 2.3.0, CUDA, and PyG, running on a system with an Intel Core i7-14700K CPU (3.40 GHz), a NVIDIA GeForce RTX 4090 GPU, and 64 GB RAM.

For \textbf{the dataset}, we used the ToN\_IoT dataset \cite{ton_iot_ref}, which contains realistic network traffic from a variety of IoT and Industrial IoT (IIoT) devices. This dataset is a suitable proxy for IoE traffic because it includes data from sensors, actuators, and gateways that exhibit communication patterns (such as periodic beacons, event-driven bursts) similar to those of smart meters and other IoE components. It also includes labeled benign and malicious traffic, enabling supervised training and evaluation. Following the setup in \cite{app14166932}, we used selected features to construct the network graphs. 80\% of the data was used for training and 20\% for testing.

For \textbf{simulating adversarial attacks}, we implemented poisoning and evasion attacks as described in \cite{10477407}. These attacks perturb the network graph by strategically adding or removing edges and altering node features to degrade model performance. We evaluated the models under perturbation rates of 0\% (no attack), 10\%, and 50\%.

For \textbf{the compared models}, we compared our GSL-enhanced models against several baselines to demonstrate their superior resilience:
\begin{itemize}
    \item \textbf{DNN:} A standard multi-layer perceptron, as used in \cite{lee2020deep}, which processes feature data without considering network structure.
    \item \textbf{GCN \& GraphSAGE:} Two popular GNN architectures widely used for network analysis \cite{wu2021graph, suarez2023graph, 10925363, app14166932}, which represent the state-of-the-art before GSL. In here, we follow the model parameters and settings in \cite{app14166932}.
    \item \textbf{GSL-GCN \& GSL-GraphSAGE:} Our proposed models, which integrate a GSL process with the GCN and GraphSAGE architectures, respectively. Here, the GNN model parameters and settings are the same as baseline GCN \& GraphSAGE. 
\end{itemize}

We report the average accuracy, precision, recall, and F1 score over 10 experimental runs.

\subsection{Results and Analysis}

\begin{figure*}[!t]
    \centering
    \includegraphics[width=\textwidth]{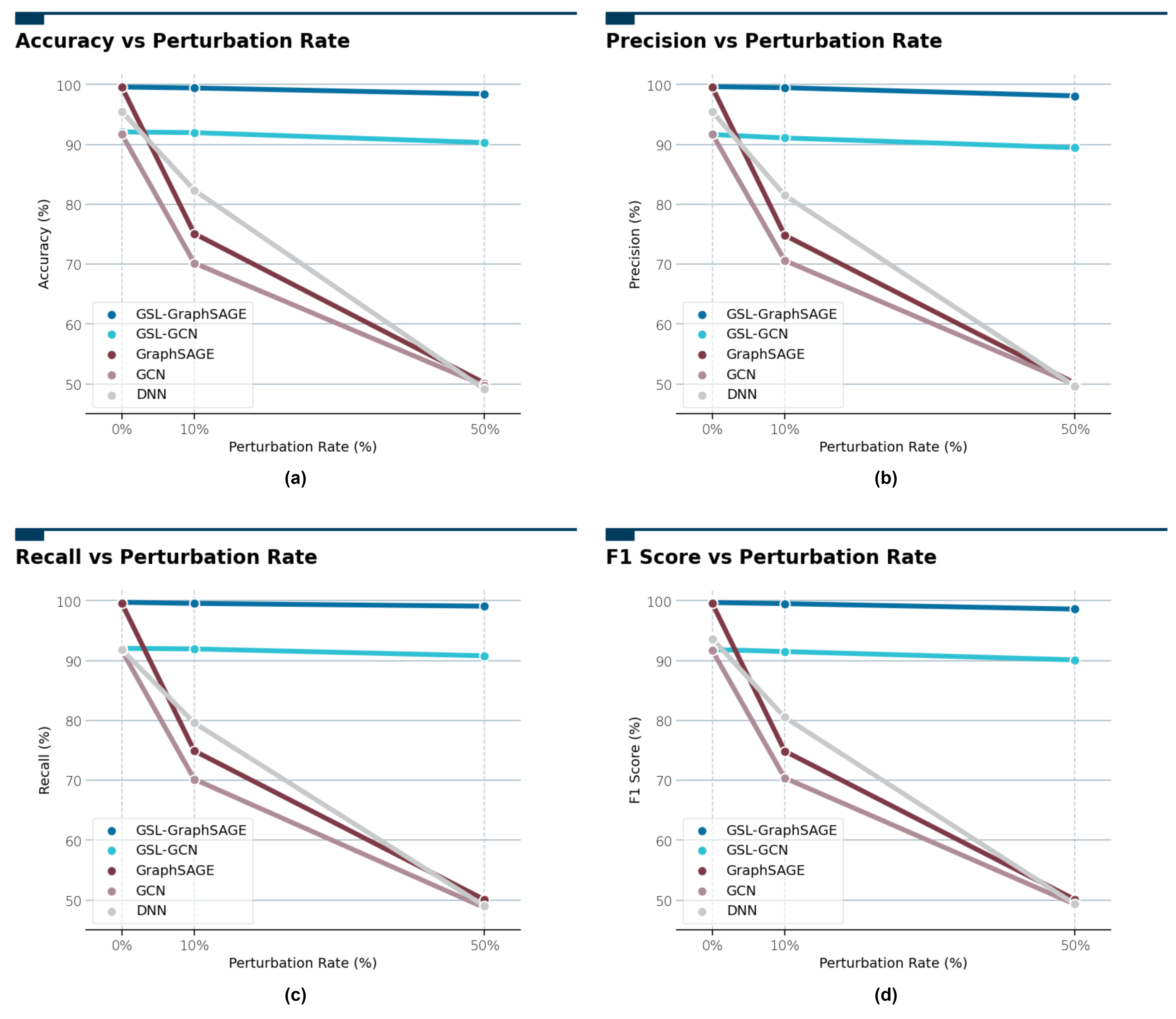}
    \caption{Evaluation of model robustness against data perturbation. The performance of GSL-GraphSAGE, GSL-GCN, GraphSAGE, GCN, and DNN is measured across varying perturbation rates (0\%, 10\%, 50\%). Subplots display results for (a) Accuracy, (b) Precision, (c) Recall, and (d) F1 Score, illustrating the superior stability of GSL-enhanced methods under perturbation.}
    \label{fig:result}
\end{figure*}

The results, presented in Figure \ref{fig:result}, unequivocally demonstrate the superior robustness of the GSL-based models. In the no-attack scenario (0\% perturbation), all graph-based models perform exceptionally well. However, as adversarial perturbations are introduced, a stark difference emerges.

At a 10\% perturbation rate, the performance of the standard DNN, GCN, and GraphSAGE models begins to decline sharply. At a 50\% rate, their accuracy, precision, recall, and F1 scores plummet to around 50-60\%, rendering them unreliable for security applications.

In stark contrast, GSL-GraphSAGE and GSL-GCN exhibit remarkable resilience. Their performance metrics remain consistently above 97\% even under the most severe 50\% perturbation. This dramatic difference in robustness is a direct result of the GSL mechanism. While standard GNNs are forced to operate on the corrupted graph, the GSL module actively identifies and prunes the spurious edges and mitigates the effects of perturbed features. By learning a cleaner, more stable graph structure, it prevents the propagation of malicious information through the network, allowing the GNN to make accurate predictions despite the adversarial noise. This confirms that integrating GSL provides a powerful defense against structural attacks, validating our proposed framework as a highly effective safeguard for IoE networks.

\section{Challenges and Future Research Directions}
\label{sec:future}
While GSL shows great promise, deploying it effectively in real-world IoE systems requires addressing several challenges and exploring new research avenues.

\subsection{Scalability} 
IoE deployments can involve millions of devices. Applying GSL to such massive graphs is computationally challenging. Future work should focus on developing distributed GSL algorithms, hierarchical graph processing, and efficient graph sampling techniques that can scale to national-level grids without compromising performance.

\subsection{Real-Time Processing} 
Security in the IoE requires rapid detection and response. The iterative nature of GSL can be time-consuming. Research into lightweight GSL architectures, hardware acceleration, and online learning algorithms that can incrementally update the graph structure is needed to meet the low-latency demands of critical infrastructure protection.

\subsection{Integration and Adaptability} 
Integrating GSL into diverse legacy systems is a practical hurdle. Furthermore, as the threat landscape evolves, security systems must adapt. A highly promising future direction is the combination of GSL with Deep Reinforcement Learning (DRL). Inspired by recent work on GNN+DRL \cite{li2025achieving}, a DRL agent could learn to dynamically adjust GSL parameters or select different structural priors in response to changing network conditions or new attack types, creating a truly adaptive and intelligent defense.

\subsection{Privacy Preservation} GSL frameworks require access to potentially sensitive network traffic data. To ensure compliance with regulations and gain user trust, research into privacy-preserving GSL is crucial. Techniques like federated learning, where models are trained locally on devices, and differential privacy, which adds statistical noise to obscure individual data points, must be adapted for graph-structured data.

\subsection{Expanding Scope and Validation} 
The current study focuses on anomaly detection. Future work should validate the GSL approach on more complex IoE simulators and extend its application to other networking tasks, such as resilient routing, resource allocation, and fault localization. Investigating the robustness of different GSL variants against an even wider range of sophisticated graph-based attacks is also a critical next step.

\section{Conclusion} \label{sec:conclusion}

Integrating IoT technologies into energy systems has revolutionized the energy sector, giving rise to the Internet of Energy (IoE). However, this evolution introduces new security challenges, particularly at the networking level. Traditional security measures often fall short in addressing sophisticated adversarial threats.
This article presented a GSL-based safeguard framework designed to enhance the resilience of IoE networks. The framework detects and mitigates adversarial attacks by jointly learning network structures and node representations. Experimental evaluations using an established dataset demonstrated its superior performance and robustness compared to conventional approaches.
As IoE evolves, incorporating advanced, adaptive security mechanisms like the proposed GSL framework will be pivotal in safeguarding critical energy infrastructures against emerging cyber threats. Future efforts must prioritize scalability, real-time performance, seamless integration, adaptability, privacy preservation, and robustness to fully secure the next generation of interconnected energy systems.


\section*{Acknowledgment}
{Zigbee®} is a registered trademark of the Connectivity Standards Alliance (CSA).  
{NB-IoT$^{TM}$} is a technology standardized by the 3rd Generation Partnership Project (3GPP).  
{LoRaWAN®} is a registered trademark of the LoRa Alliance.  
The logos used in this paper are solely for illustrative and educational purposes to reference the respective technologies. No endorsement or affiliation is implied. All rights to the trademarks and logos remain with their respective owners.
All icons from Flaticon.com have an authorized Premium license.
Guan-Yan Yang is grateful to the NSTC in Taiwan for the graduate research fellowship (NSTC-GRF) and to Professor Hung-Yi Lee for co-hosting his Ph.D research project. He is also thankful to the Norman and Lina Chang Foundation, USA, for their research scholarship. Jui-Ning Chen expresses gratitude to Dr. Lun-Wei Ku at Academia Sinica, Taiwan, for hiring her as a research assistant and providing valuable research support.

\bibliographystyle{IEEEtran}
\bibliography{IEEEabrv,ref}
%



\newpage
\begin{IEEEbiographynophoto}{Guan-Yan Yang} (Graduate Student Member, IEEE) 
    received a Bachelor's degree from the Department of Information Management at National Dong Hwa University, Hualien, Taiwan, in 2022. 
    He is currently pursuing a Ph.D. in the Department of Electrical Engineering at National Taiwan University, Taipei, Taiwan.
    In 2023, he worked as a Software Engineer at the Design Technology Platform in the Research and Development division of the Taiwan Semiconductor Manufacturing Company. Since 2024, he has been a researcher at the Taiwan Academic Cybersecurity Center and the Institute of Information Science at Academia Sinica in Taiwan. In 2024, he was awarded a scholarship from the Norman and Lina Chang Foundation in the USA.
    His research interests include security, safety, deep learning, generative AI, the Internet of Things, formal verification, and software testing. He is a member of the IEEE Computer Society, the IEEE Reliability Society, and the IEEE Consumer Technology Society.
\end{IEEEbiographynophoto}

\vspace*{-8pt}
\begin{IEEEbiographynophoto}{Jui-Ning Chen}
served as a Research Assistant at the Institute of Information Science, Academia Sinica, Taipei, Taiwan. She received the B.S. degree in Computer Science from the University of Taipei in 2022 and the M.S. degree in Electrical Engineering from National Taiwan University in 2024. 
Her research interests include natural language processing, deep learning, and fake news. Her recent research on enhancing fake news explanation with large language models was presented at NAACL 2024.
\end{IEEEbiographynophoto}
\vspace*{-8pt}
\begin{IEEEbiographynophoto}{Farn Wang} (Member, IEEE) 
    is a Full Professor at the Department of Electrical Engineering, National Taiwan University.
    He received the B.S. degree in Electrical Engineering from National Taiwan University in 1982 and the M.S. degree in Computer Engineering from National Chiao-Tung University in 1984. 
    He completed his Ph.D. in Computer Science at the University of Texas at Austin in 1993. 
    He is a founding member and chairman of the Steering Committee of the International Symposium on Automated Technology for Verification and Analysis (ATVA) from 2003 to 2022, and has served on the ATVA advisory committee since 2022.  
    He was also an Associate Editor of FMSD (International Journal on Formal Methods in System Design), Springer-Verlag.   
    His research interests include formal verification, model-checking, software testing, security, verification automation, AI, and language models.
    He has been named a World's Top 2\% Scientists in the career-long list by Stanford University since 2020.
\end{IEEEbiographynophoto}
\vspace*{-8pt}

\begin{IEEEbiographynophoto}{Kuo-Hui Yeh} (Senior Member, IEEE) 
serves as a professor at the Institute of Artificial Intelligence Innovation, National Yang Ming Chiao Tung University, Hsinchu, Taiwan. 
Prior to this appointment, he was a professor in the Department of Information Management at National Dong Hwa University, Hualien, Taiwan, from February 2012 to January 2024. 
Dr. Yeh earned his M.S. and Ph.D. degrees in Information Management from the National Taiwan University of Science and Technology, Taipei, Taiwan, in 2005 and 2010, respectively. 
He has contributed over 150 articles to esteemed journals and conferences, covering a wide array of research interests such as IoT security, Blockchain, NFC/RFID security, authentication, digital signatures, data privacy and network security. 
Furthermore, Dr. Yeh plays a pivotal role in the academic community, serving as an Associate Editor (or Editorial Board Member) for several journals, including the Journal of Information Security and Applications (JISA), Human-centric Computing and Information Sciences (HCIS), Symmetry, Journal of Internet Technology (JIT) and CMC-Computers, Materials \& Continua. 
In the professional realm, Dr. Yeh is recognized as a Senior Member of IEEE and holds memberships with ISC2, ISA, ISACA, CAA, and CCISA. His professional qualifications include certifications like CISSP, CISM, Security+, ISO 27001/27701/42001 Lead Auditor, IEC 62443-2-1 Lead Auditor, and ISA/IEC 62443 Cybersecurity Expert, covering fundamentals, risk assessment, design, and maintenance specialties. 
\end{IEEEbiographynophoto}

\end{document}